\documentclass[aps,prd,reprint]{revtex4-2}

\usepackage[utf8]{inputenc}
\usepackage{amsmath}
\usepackage{amsfonts}
\usepackage{amssymb}
\usepackage{hyperref}
\usepackage{xcolor}

\usepackage{enumitem}

\begin{document}
 
\title{Spherical symmetric fields on torsion-free Palatini Gauss-Bonnet theory}

\author{Máximo Bañados$^{(a)}$ and Daniela Bennett$^{(b)}$  \\
{\it $^{(a)}$Facultad de Física, Pontificia Universidad Católica de Chile,\\
Avenida Vicuña Mackenna 4860, Santiago, Chile \\
$^{(b)}$Grupo de Cosmología, Departamento de Física, FCFM,
Universidad de Chile, Blanco Encalada 2008, Santiago, Chile }}

\begin{abstract}
The Gauss-Bonnet density `a la Palatini' is not a total derivative in four dimensions. We study spherically symmetric fields for the torsion-free theory. The resulting equations are highly complicated but we show the existence of unexpected hidden gauge symmetries, beyond diffeomorphisms and Weyl transformations.     
\end{abstract}  
  
\maketitle

\section{Introduction}

Higher order curvature gravitational Lagrangians continue to be under intense scrutiny. In particular the Lovelock series \cite{Lovelock:1971yv} has spanned a large amount of literature due the  attractive feature of having  second order field equations.  The Lovelock series, however, is relevant only at dimensions greater than 4.  Alongside developments in higher-curvature gravity, a comparatively less studied generalization is provided by the Palatini formulation, where the relation between metric and connection is determined dynamically. While for the Einstein–Hilbert action this formulation is equivalent to the metric approach and leads back to the Levi-Civita connection, this equivalence breaks down for more general Lagrangians. Notably, the Gauss–Bonnet density, when treated à la Palatini, ceases to be a total derivative and instead yields non-trivial equations of motion for the gravitational variables.

We shall call the ``Palatini Gauss-Bonnet" action the functional,
\begin{eqnarray}
I[g,\Gamma] = k\int d^4x \sqrt{g}\, \delta^{[\mu_1 \mu_2 \mu_3\mu_4]}_{[\nu_1 \nu_2 \nu_3\nu_4]}  R^{\nu_1\nu_2}_{\ \ \mu_1 \mu_2}R^{\nu_3\nu_4}_{\ \ \mu_3 \mu_4} \label{action0}
\end{eqnarray} 
where $k$ is a dimensionless coupling, and
\begin{eqnarray}
R^{\mu \nu}_{\ \ \alpha\beta} =g^{\nu\nu'}  R^{\mu }_{\ \nu' \alpha\beta}(\Gamma).
\end{eqnarray} 
Here, $ R^{\mu }_{\ \nu' \alpha\beta}(\Gamma)$ is a function of the connection only. We use the Misner-Thorne-Wheeler convention for the curvature where the first index up, and all others down depends only on the connection. In this paper, the connection is assumed symmetric (no torsion) but otherwise arbitrary. The dynamical fields are the metric and the connection which are varied independently. 

The equations of motion are:
\begin{eqnarray}
T_{\mu\nu} \equiv  {\delta I[g,\Gamma ] \over \delta g^{\mu\nu}}&=& 0, \\ 
S^{\mu\nu}_{\ \ \ \rho} \equiv  {\delta I[g,\Gamma ] \over \delta \Gamma^{\rho}_{\ \mu\nu }} &=& 0. 
\end{eqnarray} 
with
\begin{eqnarray}
T_{\sigma\rho} &=& \sqrt{|g|}\left( \epsilon_{\mu(\sigma\alpha}^{\ \ \ \ \  \beta} R^{\mu}_{\ \rho)\alpha_1\alpha_2} +{1 \over 2} g_{\sigma\rho} \epsilon_{\mu\alpha}^{ \ \ \ \nu\beta}   \right) \times  \nonumber\\
&&\ \ \ \ \ \ \ \ \   R^{\alpha}_{\ \beta\alpha_3\alpha_4}\epsilon^{ \alpha_1\alpha_2\alpha_3\alpha_4} ,   \nonumber\\
S^{\rho\nu}_{ \ \ \ \mu} &=& -\epsilon^{\alpha_1\alpha_2\alpha_3(\rho} \nabla_{\alpha_1}(\sqrt{|g|} \epsilon_{\mu\alpha}^{ \ \ \ \nu)\beta} )  R^{\alpha}_{\ \beta\alpha_2\alpha_3 } .
\end{eqnarray} 
Recall that $T_{\mu\nu}=T_{\nu\mu}$ and $S^{\mu\nu}_{\ \ \ \rho}=S^{\nu\mu}_{\ \ \ \rho} $. This is indicated by the symmetrization parenthesis (without $1/2$).  The action is Weyl and diff invariant and consequently the following Noether identities follow, 
\begin{eqnarray}
g^{\mu\nu}T_{\mu\nu} &=&0, \nonumber \\
\nabla_\mu T^{\mu}_{\ \nu}  -{1 \over 2} T^{\alpha\beta}\nabla_{\nu}g_{\alpha\beta} + \nabla_\beta\nabla_{\alpha}S^{\alpha\beta}_{\ \ \  \nu} + S^{\alpha\beta}_{\ \ \  \mu} R^{\mu}_{\ \beta\nu\alpha}&=&0.    \nonumber
\end{eqnarray} 

Some properties of Palatini Gauss-Bonnet gravity were studied in \cite{Janssen:2019doc},\cite{Janssen:2019uao},\cite{Blumenhagen:2012ma}.   Palatini gravity is particularly relevant in theories with high powers of the curvature tensor, for example, $f(R)$ theories. See \cite{Olmo:2011uz} for a review of this subject and \cite{Barausse:2007pn} for a no-go theorem.  More recently the full version for a non-symmetric connection was analyzed in \cite{Banados:2025sww}. In this paper however, we shall concentrate on the torsion-free case having a different structure.

We study here the Schwarzschild problem for torsion-free Palatini Gauss-Bonnet gravity, described by the action (\ref{action0}). This problem is considerable more complicated than its Einstein-Hilbert counterpart. The reason is that a connection compatible with SO(3) symmetry (no torsion) has 12 arbitrary functions of the radial coordinate. The equations of motion derived from (\ref{action0}) are of course non-linear and the problem quickly becomes un-tractable analytically. The main purpose of this note is not to display a ``4d Gauss-Bonnet black hole". We did not manage to solve the spherically symmetric equations in general. However, in the process, we uncover an unexpected feature that we believe is interesting in its own right: the existence of hidden gauge symmetries.  This was uncovered by noticing that the equations of motion did not fix all the free functions in the ansatz, even after carefully fixing all the apparent symmetries of the system. To have a better understanding on this phenomena we studied the reduced Lagrangian (static and with SO(3) symmetry) and check the existence of unexpected first class constraints generating these symmetries.  

This result is in strong consistency with the $GL(4)$ theory (non-symmetric connection) recently  studied in \cite{Banados:2025sww}. In that case, there are 3 extra gauge symmetries. The $GL(4)$ connection case is somehow simpler (more structure) that the torsion-free case and a fully dynamical analysis is available. For the torsion-free case, discussed here, we will not attempt a general analysis but restricts the discussion to static spherically symmetric fields.

\section{SO(3) reduction of Palatini Gauss-Boonet gravity}

We start by isolating the static fields with SO(3) symmetry. The most general static metric consistent with SO(3) is well-known, 
 \begin{equation}
 ds^2 = f(r)\Big(-dt^2 + 2 H(r) dt dr + h(r)^2dr^2 + w(r)^2 d\Omega^2\Big) \label{so3}
 \end{equation}
Since our equations are Weyl invariant, it is convenient to  pull out the component $g_{tt}$ leaving  $-1$ in its place. The functions $H,h,w$ are so far arbitrary and therefore there is no lack of generality. Due to the Weyl invariance of the action, the function $f(r)$ will play no role at all and can be set to 1. The metric (\ref{so3}) is still invariant under the residual diffeomorphisms $t\rightarrow t + \xi^{0}(r), r\rightarrow \xi^{1}(r)$. As usual, we use these symmetries to set $w(r)=r$ and $H(r)=0$. The final form for the metric ansatz is then, 
\begin{equation}\label{metricsph}
 ds^2 = -dt^2 + h(r)^2dr^2 + r^2 d\Omega^2
 \end{equation}
having only one free function. The Weyl and diff symmetry have been fixed. 

As mentioned in the introduction, an unexpected feature of the action (\ref{action0}) is the existence of hidden gauge symmetries. These symmetries act on $h(r)$ and, a consequence, this function can be fixed to any desired value, for example $h(r)=1$.  Thus, in this theory, all static spherically symmetric metrics are gauge equivalent to flat space. (This result upgrades to a more general statement in the non-symmetric connection theory, see  \cite{Banados:2025sww}.) Flat space is certainly an interesting background, but we shall see that the choice $h(r)={1 \over r}$ allows a full exact solution that will be very useful to explore the theory.   
 
Connections invariant under $SO(3)$ are less familiar but known \cite{Hohmann:2019fvf}. We restrict our attention to non-torsion spacetimes. The resulting field has 12 arbitrary functions of $r$.  It is convenient to split the SO(3) connection in two parts, background + fluctuations, 
\begin{eqnarray}
\Gamma^{\mu}_{\ \nu\alpha}=\Gamma^{\mu}_{(0)\, \nu\alpha}+\gamma^{\mu}_{\ \nu\alpha}
\end{eqnarray} 
The only non-zero components of the ``background" are,  
\begin{eqnarray}
\Gamma^{\theta}_{(0)\, \phi\phi} &=& -\sin\theta \cos\theta,  \\
\Gamma^{\phi}_{(0)\, \theta\phi} &=& \cot\theta,  
\end{eqnarray}   
(Of course $\Gamma^{\phi}_{(0)\, \phi\theta} = \Gamma^{\phi}_{(0)\,\theta\phi} $.)  This part of the SO(3)  connection does not have any free functions or coefficients, hence the name ``background". 

The second piece contains the ``dynamical" part with arbitrary functions to be determined by the field equations \cite{Hohmann:2019fvf}, 
\begin{eqnarray}
\gamma^{t}_{\ \mu\nu} &=&  \left( \begin{array}{cccc}
s_1 & s_2 & 0 & 0 \\ 
s_2 & n_1 & 0 & 0 \\ 
0 & 0 & s_3 & 0 \\ 
0 & 0 & 0 & s_3 \sin^2\theta
\end{array} \right)  \\
\gamma^{r}_{\ \mu\nu} &=&  \left( \begin{array}{cccc}
s_4 & s_5 & 0 & 0 \\ 
s_5 & n_2 & 0 & 0 \\ 
0 & 0 & s_6 & 0 \\ 
0 & 0 & 0 & s_6 \sin^2\theta
\end{array}  \right)  \\
\gamma^{\theta}_{\ \mu\nu} &=&  \left( \begin{array}{cccc}
0 & 0 & s_7  & -s_8 \sin\theta  \\ 
0 & 0 & s_9 &  -s_{10}\sin\theta \\ 
s_7 & s_9 & 0 & 0 \\ 
-s_8\sin\theta & -s_{10}\sin\theta & 0 & 0
\end{array}  \right) \\
\gamma^{\phi}_{\ \mu\nu} &=&  \left( \begin{array}{cccc}
0 & 0 & s_8\csc\theta  & s_7   \\ 
0 & 0 & s_{10} \csc\theta &  s_{9} \\ 
s_8\csc\theta & s_{10}\csc\theta & 0 & 0 \\ 
s_7 & s_{9} & 0 & 0
\end{array}  \right) \label{connsph} 
\end{eqnarray} 
The 12 coefficients, $n_1(r),n_2(r),s_1(r),s_2(r),...,s_{10}(r)$ are all arbitrary functions of $r$. As shown in \cite{Hohmann:2019fvf}, the connection $\Gamma^{\mu}_{\ \nu\alpha}=\Gamma^{\mu}_{(0)\, \nu\alpha}+\gamma^{\mu}_{\ \nu\alpha}$ satisfies the symmetry requirement, the ``Lie Derivative" for connections,
\begin{equation}
\nabla_\alpha\nabla_\beta X^\mu + R^{\mu}_{\ \beta \sigma\alpha} X^\sigma =0.
\end{equation}
where $X^\mu$ is any of the three SO(3) Killing vectors. The fluctuation, being the difference of two connections, transforms as a tensor. This means that it satisfies the usual condition for tensors, 
\begin{equation}
{\cal L}_X \, \gamma^{\mu}_{\ \nu\alpha } =0.
\end{equation}

Before proceeding, we explain the names given to the various coefficients. The functions $\gamma^{t}_{\ rr}\equiv n_1$ and  $\gamma^{r}_{\ rr}\equiv n_2$  will appear as Lagrange multipliers enforcing 2 primary constraints. The variation with respect to the metric component $g_{rr}\equiv h^2$ gives a third primary constraint.  In this way, all $rr$ components play the role of Lagrange multipliers in  radial-quantization. On the other hand, the 10 functions $s_i(r)$ $(i=1,2,...,10)$ will satisfy first order ``dynamical" equations. Recall that the categorization ``dynamical" and ``constraints" refers, respectively, to equations with and without radial derivatives.

Our main goal in this paper is to observe that Palatini Gauss-Bonnet gravity has extra hidden gauge symmetries, independent from diff and Weyl invariance. This will be done exploring the Hamiltonian structure and prove the existence of first class constraints generating those symmetries. 

\section{Gauss-Bonnet with spherically symmetry as a non-canonical constrained system}

A useful property of spherically symmetric fields is that the reduced action give rise to the correct equations. That is, varying the full action and then impose spherical symmetry commutes with imposing spherical symmetry first and then vary with respect to the remaining fields.  This a non-trivial statement, see \cite{Torre:2010xa} for a discussion. The Palatini Gauss-Bonnet action respects this property and we can study the theory directly on a simpler Lagrangian for a reduced number of fields depending only on the coordinate $r$. 

The reduced action is a one-dimensional integral  
\begin{equation}
I[q_i,\lambda^\alpha] = \int dr\, L(q_i,q'_i,\lambda^\alpha),
\end{equation}
where $\lambda^\alpha$ are three functions built from $n_1,n_2,h$ (see below, Eq. (\ref{lm})) and  enter the Lagrangian with no derivatives. We will treat this Lagrangian as a ``dynamical"  problem where initial conditions are prescribed at some fixed radius, $r=r_0$. 

\subsection{General aspects of a non-canonical constrained system}
\label{NonCan}

Before going to the specifics of Palatini Gauss-Bonnet gravity, we review some general properties of constrained {\it non-canonical} first order systems. In our problem, both the Hamiltonian and constraints depend explicitly on $r$, which plays the role of ``time". This dependence induces extra terms in several formulas that need to be analyzed with care. 

Consider the class of Lagrangians, 
\begin{eqnarray}
L(q^a,q'^a,\lambda^\alpha,\mu^i,r) &=&  \ell_{a}(q,r)\, q\,'^{a} -H . \label{L}
\end{eqnarray}
where the total Hamiltonian $H$, 
\begin{eqnarray}
H = H_0(q,r) + \phi_\alpha(q,r)\, \lambda^\alpha +\chi_i(q,r)\,\mu^i,
\end{eqnarray} 
has a non-zero piece, $H_0(q)$, a set of first class constraints, $\phi_\alpha(q)$, and a set of second class constraints, $\chi_i(q)$. $\lambda^\alpha$ and $\mu^i$ are Lagrange multipliers. We use primes for the ``velocities" to remind that in our application $r$ plays the role of time.

The ``dynamical" fields (in the radial sense) are $q^a(r)$, with $a=1,2,...N$. For Gauss-Bonnet $N=10$ ($N$ should be even). The full set $\{q^a\}$ defines the ``phase space" of the theory that will have a Poisson bracket structure. 

The symplectic potential $\ell_{a}(q,r)$ depends on the dynamical fields $q^a(r)$ and $r$ independently. Its main property is that the 2-form, 
\begin{equation}
\omega_{ab}(q,r) = \partial_a \ell_b - \partial_b \ell_a
\end{equation}
is invertible. (If $\omega_{ab}$ was not invertible there would be more constraints.) The inverse, 
\begin{equation}
J^{ab}\omega_{bc} = \delta^a_c 
\end{equation}
defines a Poisson bracket for all phase space functions $A(q),B(q)$,
\begin{equation}\label{PB}
[A,B] \equiv \partial_a A\, J^{ab} \, \partial_ b B .
\end{equation}
This definition has all the properties of a Poisson bracket. In particular, the closure of $\omega_{ab}$ implies the Jacobi identity for $J^{ab} $. Note in particular the basic Poisson bracket,
\begin{equation}
[q^a,q^b] = J^{ab}(q)  
\end{equation}
is non-canonical because the matrix $J^{ab}(q)$ may depend on the phase space variables.  See, for example, \cite{Faddeev:1988qp} for more details on these systems. 

The Hamiltonian $H_{0}(q,r)$ is a  non-zero function of the dynamical variables.  This function will satisfy some consistency conditions in the presence of gauge invariance (see below).

The functions $\lambda^\alpha$ and $\mu^i$ are Lagrange multipliers enforcing the constraints $\phi_\alpha(q,r)=0$ and $\chi_i(q,r)=0$. We have split the constraints in two classes. The set $\phi_\alpha$ satisfy a closed algebra and are ``first class" in Dirac's terminology. That means the commute weakly with themselves, with all other constraints. In practice there exists functions  $ f^{\gamma}_{\ \alpha\beta},D^{\beta}_{(1) \alpha i }$ and $D^{j}_{(2) \alpha i }$ such that,
\begin{eqnarray}\label{class}
~[\phi_\alpha,\phi_\beta] &=& f^{\gamma}_{\ \alpha\beta}\phi_\gamma, \\ 
~[\phi_\alpha,\chi_i]&=& D^{\beta}_{(1) \alpha i }\phi_\beta + D^{j}_{(2) \alpha i }\chi_j
\end{eqnarray}
The constraints $\chi_i$ on the other hand satisfy
\begin{eqnarray}
[\chi_i,\chi_j]=C_{ij}
\end{eqnarray} 
where the matrix $C_{\ij}$ is invertible. In the Gauss-Bonnet theory there will be both first and second class constraints. 

Further conditions on the system emerge when analyzing the equations of motion and their consistency beyond the initial surface. Varying (\ref{L}) with respect to $q^a,\lambda^\alpha,\mu^i$ one finds the equations of  motion, 
\begin{eqnarray}
q'^a &=& [q^a,H] + J^{ab} \partial_r \ell_b  \nonumber \label{d1} \\
\phi_{\alpha}&=& 0 \nonumber\\
\chi_{i}&=& 0 
\end{eqnarray} 

Note the last term in (\ref{d1}), appearing due to the dependence on $r$ of the symplectic potential.

The constraints must be preserved by the evolution in $r$. In other words, one needs to check that their radial derivative is zero on the constraint surface. 

The analysis is simpler for the second class constraints. The on-shell radial derivative of $\chi_i$ is 
\begin{eqnarray}\label{cons1}
0={d\chi_i \over dr } &=&  [\chi_i, H_0] + [\chi_i,\phi_\beta]\lambda^\beta+ [\chi_i,\chi_j]\mu^j +   \nonumber\\
&& \ + {\partial \chi_\alpha \over \partial r } + \partial_a \chi_\alpha J^{ab} \partial_r \ell_b
\end{eqnarray} 
The crucial point here is that, by definition of second class, $[\chi_i,\chi_j]=C_{ij}$ is invertible. Then, (\ref{cons1}) merely fixes the Lagrange multipliers $\mu^j$ in terms of the other variables. 

Consistency of first class constraints is very different. The on-shell radial derivative of $\phi_\alpha$ is 
\begin{eqnarray}
0={d\phi_\alpha \over dr } &=&  [\phi_\alpha, H_0] + [\phi_\alpha,\phi_\beta]\lambda^\beta+ [\phi_\alpha,\chi_i]\mu^i +   \nonumber\\
&& \ + {\partial \phi_\alpha \over \partial r } + \partial_a \phi_\alpha J^{ab} \partial_r \ell_b\nonumber1 \\
&\approx &   [\phi_\alpha, H_0]  + {\partial \phi_\alpha \over \partial r } + \partial_a \phi_\alpha J^{ab} \partial_r \ell_b  \label{cons2}
\end{eqnarray} 
In the second line we have used that the brackets $[\phi_\alpha,\phi_\beta]$ and $[\phi_\alpha,\chi_i]$ are weakly zero. (If the Hamiltonian and symplectic potential did not depend on the evolution variable, $r$ in or case, this condition would imply the well-known condition $ [\phi_\alpha, H_0]\approx 0$.)  

Condition (\ref{cons2}) can give two different possibilities:  It may represent a new constraint, which would need to be added to the system and then check its own consistency with radial evolution. This is called a secondary constraint. A second possibility is that the Hamiltonian $H_0$, the constraints $\phi_\alpha$ and symplectic potential $\ell_a$ has been chosen such that (\ref{cons2}) is satisfied automatically (weakly).  

If all constraints have been found, the only possibility is that (\ref{cons2}) is an identity (weak). This means there must exists functions $ F^{\alpha}_{\ \beta}$ and $S^{i}_{\ \alpha}$ such that
\begin{equation}
 [\phi_\alpha, H_0] + {\partial \phi_\alpha \over \partial r }+ \partial_a \phi_\alpha J^{ab} \partial_r \ell_b =\phi_\alpha \, F^{\alpha}_{\ \beta} + \chi_i S^{i}_{\ \alpha} \label{3}
\end{equation} 
It is good exercise, to check the consistency of these models working out these identities finding the functions $F$ and $S$. We have done this for the Palatini Gauss-Bonnet theory (although  expressions are too long to be displayed explicitly). The general structure will be discussed in Sec. \ref{Linear}. 

If (\ref{3}) is satisfied, and $F^{\alpha}_{\ \beta},\ S^{i}_{\ \beta} $ are known, then $\phi_\alpha$ generates the gauge symmetry,
\begin{eqnarray}
\delta q^a &=& [q^a , \phi_\alpha]\epsilon^\alpha(r) \nonumber \\
\delta \lambda^\alpha &=& \epsilon'^\alpha(r) + f^{\alpha}_{\ \beta\gamma}\lambda^\beta \epsilon^\gamma  + F^{\alpha}_{\ \beta}  \epsilon^\beta \nonumber \\
\delta \mu^i &=& S^{i}_{\ \beta}  \epsilon^\beta(r)   \label{gauge}
\end{eqnarray} 
with $\epsilon^\alpha(r)$ an {\it arbitrary} function of $r$. The proof that this transformation leave the action invariant is straightforward.  Observe that the coefficients $F^{\alpha}_{\ \beta}$ and $S^{i}_{\ \beta} $ entering in (\ref{3}) appear in the transformation.     

These symmetries implies that the equations of motion do not fix all dynamical variables. For each gauge symmetry there will be one function that remains arbitrary and must be fixed via a gauge condition.  In the Palatini Gauss-Bonnet theory we shall find two such functions (in the extended formalism, to be discussed below).

\subsection{Application to Palatini Gauss-Bonnet with spherical symmetry}

Palatini Gauss-Bonnet gravity for spherically symmetric fields provides an example of the structure just discussed. All ingredients show up: a non-canonical phase space with a Poisson bracket, primary and secondary constraints, first class and second class constraints. The explicit formulas are long and complicated. But, with the help of algebraic computing the main equations can be tamed to reach clear conclusions. 

We plug the metric (\ref{metricsph}) and connection (\ref{connsph}) into the Lagrangian (\ref{action0}) obtaining a reduced Lagrangian for the 13 functions $h(r),n_1(r),n_2(r),s_i(r)$. Before displaying the result we make the following (invertible) redefinition of fields ($s_i \rightarrow q_i$):
\begin{eqnarray}
s_{1} &=& q_1 \nonumber\\
s_2 &=& h\, q_2 \nonumber\\
s_3 &=& q_{3} \nonumber\\
s_4 &=& {1 \over h}\, q_4 \nonumber\\
s_5 &=& q_5 \nonumber\\
s_6 &=& {1 \over h}\, q_6 \nonumber\\
s_7 &=& q_7 \nonumber\\
s_8 &=& q_9 \nonumber\\
s_9 &=& h\, q_9 \nonumber\\
s_{10} &=& h\, q_{10} 
\end{eqnarray} 
Similarly, we replace $n_1,n_2,h$ by three new functions, $\lambda_1,\lambda_2,\lambda_3$ via
\begin{eqnarray}
h &=& \lambda_3,\nonumber\\
n_1 &=& \lambda_1\lambda_3, \nonumber\\
n_2 &=& \lambda_2 +  {\lambda_3' \over \lambda_3}.  \label{lm}
\end{eqnarray} 
After discarding a boundary term, the Gauss-Bonnet Lagrangian written in these variables has the form of a constrained non-canonical system, 
\begin{equation}
L(q,\lambda^\alpha) = \ell_{a}(q,r)\, q'^a - \lambda^\alpha \psi_\alpha(q,r) \label{redl0}
\end{equation}
where $a=1,2,...,10$ and $\alpha=1,2,3$. The symplectic potential functions are the  following, 
\begin{eqnarray}
\ell_1 &=&  0 \nonumber\\
\ell_2 &=& -r^2 (q _3q_7 + q_6 q_9)\nonumber\\ 
\ell_3 &=& -r^2 ( q_3q_4+q_2 q_7 - q_6 q_7 - q_7 q_9 + q_5q_6+q_5 q_9 + q_8 q_{10})\nonumber\\
\ell_4 &=& -r^2( q_3 q_7 + q_6 q_9)\nonumber\\
\ell_5 &=& 0\nonumber\\
\ell_6 &=& r^2( q_2 q_6 + q_1q_3-q_1q_7 - q_3q_7 + q_7^2 - q_8^2 -q_4 q_9)\nonumber\\
\ell_7 &=& -r^2(q_3 q_4 + q_2 q_7 - q_6 q_7- q_7q_9 + q_5q_6+q_5q_9 + q_8 q_{10}  )\nonumber\\
\ell_8 &=& -r^2( q_2 q_8 + 3 q_6 q_8 - q_8 q_9 + 2q_3 q_{10} + q_5 q_{10} - q_7 q_{10})\nonumber\\
\ell_9 &=& r^2(-q_2 q_6 + q_3 q_7 - q_7^2 + q_1(-q_3+q_7) + q_8^2 + q_4 q_9 )\nonumber\\
\ell_{10} &=& r^2(q_1q_8 - q_3 q_8 - 2 q_7 q_8 + q_4 q_{10}) \nonumber
\end{eqnarray} 
and the constraints, 
\begin{eqnarray}
\psi_1 &=& q_1-q_5 + r^2(q_7{}^3-q_1 q_7{}^2- q_3 q_7{}^2+q_6{}^2    q_7+q_8{}^2 q_7 \nonumber\\ &&  +2 q_1 q_3 q_7-q_3 q_5 
   q_7-q_4 q_9 q_7+q_6 q_9 q_7-q_5
   q_6{}^2  \nonumber\\ &&  -q_1 q_8{}^2+q_3 q_8{}^2-q_3 q_4
  q_6+q_1 q_6 q_9-2 q_5 q_6 q_9-q_4 q_8  \nonumber\\ && 
   q_{10}-q_6 q_8 q_{10})  \nonumber\\
\psi_2 &=&     q_2-q_4 + r^2( q_2 q_6{}^2+q_7{}^2 q_6-q_8{}^2  q_6+q_1 q_3 \nonumber\\ && 
   q_6-q_1 q_7 q_6-q_3 q_7 q_6+2 q_2 q_9
   q_6-2 q_4 q_9 q_6-q_4 q_9{}^2 \nonumber\\ &&  -q_4
   q_{10}{}^2+q_2 q_3 q_7-q_3 q_4 q_7+q_7{}^2
   q_9-q_8{}^2 q_9+q_1 q_3 q_9\nonumber\\ &&  -q_1 q_7
   q_9-q_3 q_7 q_9-q_1 q_8 q_{10}+q_3 q_8
   q_{10}+2 q_7 q_8 q_{10}) \nonumber\\
\psi_3 &=&   -q_2^2+q_4 q_2+q_5^2-q_1 q_5 + \nonumber\\ 
   && r^2(q_4 q_9{}^3+q_5{}^2 q_9{}^2+2 q_8{}^2 q_9{}^2-q_1 q_3 q_9{}^2+q_3 q_5 q_9{}^2 \nonumber\\ 
   &&  -q_2 q_6 q_9{}^2+q_4 q_6 q_9{}^2+q_1 q_7 q_9{}^2-  2 q_5 q_7 q_9{}^2 \nonumber\\ 
   && -q_2 q_6{}^2 q_9-2 q_2 q_7{}^2 q_9-2 q_2 q_8{}^2 q_9+q_4 q_{10}{}^2 q_9+q_3{}^2 q_4 \nonumber\\ 
   &&    q_9-q_2 q_3 q_5 q_9-q_2{}^2 q_6 q_9+2q_5{}^2 q_6 q_9-q_1 q_3 q_6 q_9\nonumber\\ 
   && +q_2 q_4+q_6 q_9-q_1 q_5 q_6 q_9+q_3 q_5 q_6 q_9+2 q_2 q_3 q_7 q_9 \nonumber\\ 
   && -q_3 q_4 q_7 q_9 +2 q_2 q_5 q_7 q_9+q_1 q_6 q_7 q_9  -2 q_5 q_6 q_7 q_9 \nonumber\\ 
   && +2 q_1 q_8 q_{10} q_9-4 q_7 q_8 q_{10} q_9  +q_2{}^2 q_7{}^2+ q_3 q_5 q_7{}^2 \nonumber\\ 
   && - q_2 q_6 q_7{}^2+q_2{}^2 q_8{}^2-q_3 q_5 q_8{}^2+ q_2 q_6  q_8{}^2+\nonumber\\ 
   &&q_5{}^2 q_{10}{}^2+2 q_7{}^2 q_{10}{}^2+q_1q_3 q_{10}{}^2-q_3 q_5 q_{10}{}^2+ \nonumber\\ 
   && q_2 q_6    q_{10}{}^2-q_4 q_6 q_{10}{}^2 -q_1 q_7 q_{10}{}^2-2 q_5 q_7 q_{10}{}^2 \nonumber\\ 
   &&-q_2 q_3{}^2 q_4+q_1 q_3{}^2 q_5+ q_3 q_5{}^2 q_7-2 q_2{}^2 q_3 q_7 \nonumber\\ 
   &&+2 q_2 q_3 q_4 q_7 -q_3{}^2 q_5 q_7-2 q_1 q_3 q_5 q_7   +q_2 q_3 q_6 q_7\nonumber\\ 
   &&+q_2 q_5 q_6 q_7-2 q_2 q_3 q_8 q_{10}+q_3 q_4 q_8 q_{10}  \nonumber\\ 
   &&+2 q_2 q_5 q_8 q_{10} -q_1 q_6 q_8 q_{10}+2 q_5 q_6 q_8 q_{10})\nonumber
\end{eqnarray}

By direct calculation one can check that the associated symplectic form $w_{ab}$ is invertible, although it depends on the canonical variables ($q^a$) and is degenerate at isolated surfaces. We shall stay at the stable regions with maximum rank. The Poisson bracket is defined as in (\ref{PB}), and the equations of motion take the Hamilton form,
\begin{eqnarray}
q'^a &=&  [q^a, H] + J^{ab}  {\partial l_b \over \partial r} .    \label{eq1} \\
\psi_\alpha &=& 0\label{const1}
\end{eqnarray} 
where the Hamiltonian is purely a combination of the constraints, 
\begin{eqnarray}
H = \lambda^\alpha \psi_\alpha. 
\end{eqnarray} 
This means that the radial evolution is driven purely by the constraints. This will change when analyzing the system perturbatively. The system is not yet exactly in the form (\ref{L}) because the $\psi_\alpha$ are not separated in  first and second class constraints. 

The above formulas show that the Schwarzschild problem in torsion-free Palatini Gauss-Bonnet gravity is far more complicated than the Einstein-Hilbert counterpart. We will explore this system first by a series analysis. This will gives a path to find a particular background. Then, we study fluctuations on that background confirming the existence of a secondary constraint and 2 gauge symmetries. 

Before going into details, a summary of results is the following. Running Dirac's consistency algorithm one secondary constraint shows up and the process stops (no tertiary constraints). So, in total, there are 4 constraints and they split as 2+2: Two first class generating two gauge symmetries, and 2 second class. Our main conclusion, torsion-free static SO(3) Palatini Gauss-Bonnet gravity has two extra gauge symmetries, beyond diffeomorphisms and Weyl transformations, hence the name ``extra hidden" symmetries. 

We shall work in the extended formalism where all constraints (primary and secondary) are added to the action with associated Lagrange multipliers. Incorporating the secondary constraint into the Hamiltonian with an arbitrary Lagrange multiplier modifies the covariant equations. Following  Dirac, however, we keep the fourth Lagrange multiplier in the understanding that the Physics is not altered. See \cite{Henneaux:1992ig} for the general analysis of the Dirac formalism, and a discussions on Dirac's conjecture.

\section{Series structure and an exact background}

Attempting to find the analytic solution to the system of equations descending from (\ref{redl0}) is worthless. We shall focus on a less ambitious goal, namely, the counting of  how many ``degrees of freedom" that is constants of integration this system has. 

In principle, there are 10 first order equations and 3 constraints.  But this is not the whole story: we need to check that the symplectic form is invertible and that  there are no secondary constraints. The first assertion is true, the symplectic matrix is invertible (even on the constraint surface) but the second is not true, there is a secondary constraint.

We shall study the system in two different related ways. We first analyze the equations of motion via a series expansion starting at an arbitrary point $r=r_1$. This analysis shows clearly the presence of a secondary constraint. And, luckily, the series expansions suggests a good ansatz and an exact background becomes available. We shall then analyze the theory by perturbing that background.

\subsection{Series solution} 

Many ingredients of a constrained Hamiltonian system are captured by looking at a series expansion of the equations. Consider the following expansion for all variables, 
\begin{eqnarray}
q_i(r) &=& \sum_{n=0}^\infty A_{i,n} (r/r_1-1)^n , \ \ \ \   ( i=1,2..,10)\\
\lambda_\alpha(r) &=& \sum_{n=0}^\infty u_{\alpha,n} (r/r_1-1)^n \ \ \ \ \ (\alpha = 1,2,3) 
\end{eqnarray} 
where $r_1\neq 0$ is some arbitrary fixed point. We choose not to expand around the origin to capture the structure of the equations without getting mix up with the problem of singular coordinates at $r=0$. We plug the series expansion into the equations and constraints and the following structure emerges (recall ``dynamical" and ``constraints" refer to equations (\ref{eq1}) with radial derivatives and  (\ref{const1}) purely algebraic restrictions, respectively).  

\begin{enumerate}[start=0]
\item Order zero.
\begin{itemize}
\item \textbf{Constraints:}  Since there are three constraints, naively, the 10 $A_{i,0}$'s should be split in 7+3. 3 being fixed by the constraints, and 7 left free.  However, this is not what happens. Only 6 $A_{i,0}$'s can be fixed to arbitrary values. There is another condition for the order zero parameters $A_{i,0}$ that will show up at order one. This is the signal of a secondary constraint. If one did fix 7 $A_{i,0}$'s, the process stops with no solutions. So, we fix only 6 $A_{i,0}$'s to arbitrary values, determine 3 by the constraints and leave 1 free.

\item \textbf{Dynamical equations: }
  The dynamical equations (\ref{eq1}), as expected, fix all coefficients $A_{i,1}$ in terms of the $A_{i,0}$.  This confirms that the symplectic two form is invertible. 
\end{itemize}

\item Order one.
\begin{itemize}
\item \textbf{Constraints:} Two constraints fix two Lagrange multipliers coefficients, for example, $u_{1,0},u_{2,0}$. This signals the presence of second class constraints. The third constraint fixes the coefficient $A_{i,0}$ left free at order zero. The third Lagrange multiplier is left free. This signals the presence of a gauge symmetry. 

\item  \textbf{Dynamical equations.} These equations fix all coefficients $A_{i,2}$ in terms of the lower ones. Again, confirming that the symplectic two form is invertible.
\end{itemize}

\item Order two and beyond, the problem is purely algorithmic: 

\begin{itemize}
\item \textbf{Constraints:} At each order two constraints fix $u_{1,n},u_{2,n}$ while the third constraint is automatically satisfied. The third Lagrange multiplier, with coefficients $u_{3,n}$,  is left free at all others.  \\

\textbf{Dynamical equations.} These equations fix all $A_{i,n+1}$ in terms of the lower ones.

\end{itemize}

\end{enumerate}

In summary, six initial conditions, $A_{i,0}$'s, can be prescribed to arbitrary values. The equations fix all others coefficients $A_{i,n}$ ($n>0$)  in terms of them. The Lagrange multipliers coefficients $u_{1,n},u_{2,n}$ are fully fixed, while $u_{3,n}$ are not fixed at all. This indicates the presence of two second class constraints plus (at least) one first class. In the next paragraph we shall display the full gauge structure of the system by linearizing around an exact background. 

As an example, let us fix 6  $A_{i,0}$'s by, 
\begin{eqnarray}
& A_{3,0}=\kappa,\ \ A_{6,0}=0,\ \ A_{7,0}=0,\ \ \\
&  A_{8,0}=k_1,\ \ A_{9,0}=1,\ \ A_{10,0}=k_1.\ \ 
\end{eqnarray}
leaving two parameters $\kappa,k_1$ free. The three constraints at order zero imply
\begin{eqnarray}
A_{1,0} &=& -k_1^2 r_1^2 \kappa + A_{5,0}, \\
A_{2,0} &=& - A_{5,0}, \\
A_{4,0} &= &k_1^2 r_1^2 \kappa - A_{5,0}, 
\end{eqnarray}
Observe that $A_{5,0}$ is still free. The dynamical equations at this order fix all coefficients $A_{i,1}$. At order one, as claimed, we find a condition for $A_{5,0}$ plus two conditions for $u_{1,0},u_{2,0}$ (the equation for $A_{5,0}$ is cubic with three solutions, we choose the simplest $A_{5,0}=k_1^2 r_1^2 \kappa$). Iterating this process all coefficients can be found except $u_{3,n}$ which are left free, confirming that  $\lambda_3$ is free. Since everything indicates  $\lambda_3$ is not fixed by the equations of motion we set it to a convenient value, 
\begin{equation}
\lambda_3 = {1 \over r}
\end{equation} 
With this choice of ``gauge" the series solution for all connection components have simple forms, 
\begin{eqnarray}
q_1(r) &=&  -\frac{1}{2} \kappa  k_1^2 (r-r_1) (r+r_1)+\cdots \nonumber \\
q_2(r) &=&  -\frac{1}{2} \kappa  k_1^2 \left(r^2+r_1^2\right)+\cdots\nonumber \\
q_3(r) &=&  \frac{\kappa  \left(-4 r^3+15 r^2 r_1-20 r r_1^2+11 r_1^3\right)}{2 r_1^3}+\cdots \nonumber \\
q_4(r) &=& \frac{1}{2} \kappa  k_1^2 (r-r_1) (r+r_1)+\cdots \nonumber\\
q_5(r) &=& \frac{1}{2} \kappa  k_1^2 \left(r^2+r_1^2\right)+\cdots \nonumber\\
q_6(r) &=&  -\frac{\kappa  (r-r_1) \left(4 r^2-11 r r_1+9 r_1^2\right)}{2 r_1^3}+\cdots \nonumber \\
q_7(r) &=& 0 +\cdots \nonumber \\
q_8(r) &=& k_1 +\cdots \nonumber \\
q_9(r) &=& 0 +\cdots \nonumber \\
q_{10}(r) &=& k_1 +\cdots \label{series}
\end{eqnarray} 
In each case the three dots indicate corrections ${\cal O}(r-r_1)^2$.

\subsection{An exact background}

Finding exact solutions that solve all equations derived from the Lagrangian (\ref{redl0}) is not an easy task.  Luckily, the series solution discussed in the last paragraph shows a path to, at least, a particular solution. Indeed,  observe that  $q_8(r)$ and $q_{10}(r)$ are constant 
$q_7(r)$ and $q_9(r)=0$ are zero. We have checked this properties to higher orders. Indeed an exact solution can be within this family. 

Let us then try the following ansatz on the full  equations, 
\begin{eqnarray}
q_7 = q_9 &=& 0 \\
q_8 = q_{10} &=& k_1 \\
\lambda_3 &=& {k_2 \over r},
\end{eqnarray} 
where $k_1,k_2$ are arbitrary constants. Nicely, the full system of equations can be solved. We skip details. The full solution is: 
\begin{eqnarray}
q_1(r) &=& k\left( -p + {r^2 \over r_0^2} \right) \nonumber  \\
q_2(r) &=& k\left( p + {r^2 \over r_0^2} \right)  \nonumber\\
q_3(r) &=& k\left( -1 - {pr_0^2 \over r^2} \right)  \nonumber\\
q_4(r) &=& k\left( p - {r^2 \over r_0^2} \right) \nonumber \\
q_5(r) &=& k\left(- p - {r^2 \over r_0^2} \right) \nonumber\\
q_6(r) &=& k\left( 1 - {pr_0^2 \over r^2} \right) \nonumber\\
q_7(r) &=& 0 \nonumber\\
q_8(r) &=& {1 \over r_0}\nonumber \\
q_9(r) &=& 0 \nonumber\\
q_{10}(r) &=& {1 \over r_0}. \label{solex1}
\end{eqnarray} 
Here the parameters $r_0,p,k$ are combinations of $k_1,k_2$ and a third integration constant that appear in the process. The Lagrange multipliers take the form, 
\begin{eqnarray}
\lambda_1(r) &=&  {1 \over 2r} \left( 1 - {r^2 \over p r_0^2}\right) \nonumber \\
\lambda_2(r) &=&  {1 \over 2r} \left( -1 + {r^2 \over p r_0^2}\right) \nonumber \\
\lambda_3(r) &=&  -{1 \over 2kp} {1 \over r} \label{solex2}
\end{eqnarray} 
The fields (\ref{solex1}) and (\ref{solex2}) provide a solution to the non-linear equations for arbitrary values of the parameters.  We shall use this background now to study the full theory (with spherical symmetry). 

\section{Linear theory} 
\label{Linear}

The existence of an exact background allows exploration of the theory by linearizing the equations of motion. This analysis will confirm the existence of an extra gauge symmetry.  Several results displayed in this section  can be upgraded to the full theory. We shall not include all expressions but only discuss the general structure of the results.  

Consider the expansion around the background described in last section. We name the exact solution by capital letters, 
\begin{eqnarray}
q_a &=& Q_a  + \epsilon z_a  \ \ \ \ \ \  (a=1,2,...10). \\
\lambda_I &=& \Lambda_I + \epsilon \rho_I \ \ \ \ \ \  (I=1,2,3).
\end{eqnarray}  
where $\epsilon$ is a small number. Here $Q_a,\Lambda_I$ are the exact solution (background) and $z_a,\rho_I$ the fluctuations. Expanding the Lagrangian (\ref{redl0}) to second order, dropping some boundary terms, we find,
\begin{equation}
L(z^a,\rho^I) = l_a(z,r) z'^a - H_0(z,r) - \rho^I \psi_I (z,r)
\end{equation} 
Since we expand to second order, $l_a$ and $\psi_I$ are linear in the variables $z^a$ while $H_0$ is quadratic, 
\begin{eqnarray}
l_a(z^a,r)  &=&  l_{ab}(r) z^b \\
\psi_I(z^a,r)  &=&  \psi_{I b}(r) z^b \\
H_0(z^a,r)  &=&  H_{0ab}(r) z^a z^b  
\end{eqnarray} 
The coefficients $l_{ab},\psi_{\alpha b},H_{0ab}$ depend on $r$ and the background quantities.  The symplectic matrix has a non-zero determinant, 
\begin{equation}
\mbox{det}(\omega) = 64 k^6 p^4 r^8 r_0^4 ( (1+p)r^2 -pr_0^2),
\end{equation}
except at the points $r=0$ and $r^2={pr_0^2\over(1+p) }$.  Since we are exploring the linear theory we shall be away from delicate points. The consistency of the  constraints, 
\begin{equation} \label{consislin}
{d\psi_I \over  dr}=0 \ \ \ \ \ \ \ \  (I=1,2,3)
\end{equation}
can now be checked easily because the equations are linear.  As anticipated, equations (\ref{consislin}) fix two Lagrange multipliers, say $\rho^1$ and $\rho^2$, but the third condition is a new algebraic equation for the dynamical variables $z^a$. This is the secondary constraint. 
We add this new constraint to the Hamiltonian and in a slight abuse of language  we carry on denoting the constraint surface as $\psi_I =0$ where now the index runs $I=1,2,3,4$. 

We run again (\ref{consislin}), with $I=1,2,3,4$, to find two conditions on the Lagrange multipliers and nothing else. That is, the process stops at this point. There are in total 4 constraints. By direct calculation we computed their Poisson brackets and find 2 first class and 2 second class. In the linear theory, the 4 constraints can easily be disentangle in two first class and 2 second class but we shall note include the formulas here because they are long and not enlightening. Suffice to say that all the general results discussed in Sec. \ref{NonCan} hold. In particular it can be checked explicitly that the 2 first class constraints do generate symmetries of the extended Lagrangian, as discussed in Eq. (\ref{gauge}).  
 
\section{Conclusions}

To conclude, we  have considered in this paper fields with spherical symmetry for the theory (\ref{action0}). Importantly this theory is treated a la Palatini, otherwise the action (\ref{action0}) is purely a boundary term. 

After removing all coordinate freedom plus Weyl invariance, we end up with a system of equations for 13 functions of $r$ that one would expect to be fixed by the field equations. The procedure so far mimics exactly what one does to find the Schwarzschild solution.  However, we have found that one of the functions is not fixed by the equations of motion signaling an extra ``hidden" gauge symmetry.  We have then study the reduced Lagrangian (with spherical symmetry) and proved that it reduces to a constrained non-canonical system. In this way the gauge structure is clear and the appearance of 2 first class constraints confirms the existence of extra gauge symmetries. These symmetries do not seem to have an obvious covariant interpretation, but this is an open question. 

There are many future lines of research. To start, the full solution to the equations and their properties. The equations are complicated, but the existence or not of an exact solution is an open problem.  Since there exists two concepts of curvature in this theory, the Levi-Civita and the purely connection one, extracting the geometrical properties of classical configurations raises new   challenges. Note that using the extra gauge symmetry one can always set the metric to  be flat, via a gauge choice. Indeed, $h=\lambda_3$ is arbitrary so we can set $h=1$ rendering the metric (\ref{metricsph}) flat, and all dynamics is pushed to the connection.  This phenomenon was also observed in \cite{Banados:2025sww} using the GL(4) symmetry (present in the non-symmetric connection theory). Being quadratic in the curvature, Palatini Gauss-Bonnet theory is expected to contribute in the large curvature regions, where this property seems worth further study. 

Having an exact background also suggest studying general perturbations to explore the theory beyond spherical symmetry. We plan to do this in the near future. Finally, the study of the full dynamics, along the lines of \cite{Banados:2025sww}. The definition and determination of the maximum rank for these theories (with a symmetric connection) is a difficult challenge. Unexpectedly, the torsion-free theory is more complicated that the full theory which has interesting similarities with Yang-Mills theory.

\section{Acknowledgements} 
 
We would like to thank I. Cruz who participated in an early stage of this project. MB would like to thank  M. Henneaux, G. Barnich and A. Faraggi for many useful conversations while this paper was being prepared, and A . Castro and A. Sfondrini for hospitality in Cambridge and Padova where part of this work was done. He would also like to thank J.M. Martín-García, L. Stein, J. Margalef and A. García-Parrado for help on the xAct system.  M.B was partially funded by a Grant from Vicerrectoría Académica UC-Chile.

\end{document}